\documentclass{article}
\normalmarginpar
\twocolumn
\makeatletter
\gdef\@proofbox{\relax}
\long\def\proofbox#1{\gdef\@proofbox{#1}}
\proofbox{\small{\sl PhysComp96}\\\ifx\UNDEF\@fullpaper
Extended abstract\else Full paper\fi\\Draft, \@date}
\gdef\fullpaper{\gdef\@fullpaper{}}
\def\affil#1{\\{\small#1\par}}
\gdef\@author{John Doe1\affil{No-Name University, Shipping Dept.}}
\long\def\author#1{\gdef\@author{#1}}
\gdef\@abstract{}
\long\def\abstract#1{\gdef\@abstract{#1}}
\def\@maketitle{\newpage\leavevmode
  \begin{minipage}[t]{0.30\textwidth}
    \hrule height0pt
    \raggedright
    \mbox{}\par
    \@proofbox
  \end{minipage}\relax
  \begin{minipage}[t]{0.70\textwidth}
    \hrule height0pt
    \raggedleft
    \LARGE\@title\par
    \vskip4pt
    \large\@author
  \end{minipage}
  \vskip8pt
  \ifx\@abstract\@empty\else{\vskip.5em\leftskip1.5in%
\parskip4pt\small\@abstract\par\vskip.5em}\fi
  \rule{\textwidth}{0.4pt}
  \vskip16pt}

\DeclareRobustCommand\em
        {\@nomath\em \ifdim \fontdimen\@ne\font >\z@
                       \upshape \else \slshape \fi}

\def\@begintheorem#1#2{\sl \trivlist \item[\hskip \labelsep{\bf #1\ #2}]}
\def\@opargbegintheorem#1#2#3{\sl \trivlist
     \item[\hskip \labelsep{\bf #1\ #2\ (#3)}]}
\newcommand{\sect}[1]{\S\ref{sect.#1}}

\newcommand{\sectlabel}[1]{\label{sect.#1}}

\def\@arabic#1{\number #1}
\makeatother

\title{Teleportation as a\\quantum computation}
\author{Gilles Brassard, {\sc frsc}\thanks{Supported in part by
{\sc Nserc} and {\sc Fcar}}
\affil{Universit\'e de Montr\'eal\thanks{D\'epartement~IRO,
C.P.~6128, succursale centre--ville, Montr\'eal, \mbox{Canada H3C 3J7}.
email: brassard@iro.umontreal.ca}}}
\date{12 May 1996}

\newcommand{\ket}[1]{\mbox{$| #1 \rangle$}}
\newcommand{\norm}[1]{\mbox{$| #1 |^2$}}
\newcommand{\state}[2]{\mbox{$(\mbox{}_{#2}^{#1})$}}
\begin{document}

\maketitle

\section{Introduction}\sectlabel{intro}

Among the many exciting new applications of quantum physics
in the realm of computation and information theory,
I~am particularly fond of quantum cryptography,
quantum computing and quantum teleportation~\cite{LNCS}.
Quantum cryptography allows for the confidential transmission
of classical information under the nose of an eavesdropper,
regardless of her computing power or technological
sophistication~\cite{BB84,BBBSS,sciam}.   Quantum computing allows for an
exponential amount of computation to take place simultaneously in a single
piece of hardware~\cite{feynman,deutsch}; 
of~particular interest is the ability of quantum computers to factorize
numbers very efficiently~\cite{shor}, with dramatic implications for
classical cryptography~\cite{RSA}.  Quantum teleportation allows for the
transmission of quantum information to a distant location despite the
impossibility of measuring or broadcasting the information to be
transmitted~\cite{BBCJPW}. Each of these concepts had a strong overtone of
science fiction when they were first introduced.

If~asked to rank these ideas on a scale of technological
difficulty, it is tempting to think that quantum cryptography
is easiest while quantum teleportation is the most outrageous---especially
when it comes to teleporting goulash~\cite{IBM}!
This ranking is correct with respect to quantum cryptography,
whose feasibility has been demonstrated by several experimental
prototypes capable of reliably transmitting confidential information
over distances of tens of kilometres~\cite{townsend,gisin,hughes}.
The situation is less clear when it comes to comparing the technological
feasibility of quantum computing with that of quantum teleportation.

On~the one hand, quantum teleportation can be implemented with a quantum
circuit that is much simpler than that required by any
nontrivial quantum computational task: the~state of an
arbitrary qubit (quantum bit) can be teleported with as few as two quantum
exclusive-or (controlled-not) gates.  Thus, quantum teleportation
is significantly easier to implement than quantum computing if we
are concerned only with the complexity of the required circuitry.

On~the other hand, quantum computing is meaningful
even if it takes place very quickly---indeed its primary purpose
is increased computational speed---and within a small region of space.
Quite the opposite,
the interest of quantum teleportation would be greatly reduced if the actual
teleportation had to take place immediately after the required preparation.
Thus, a working demonstration of quantum teleportation is likely to be seen
before the quantum factorization of even a very small integer is achieved, but
quantum teleportation across significant time and space will have to await a
technology that allows for the efficient long-term storage of quantum
information.
Nevertheless, it may be that short-distance quantum teleportation
will play a role in transporting quantum information inside quantum
computers.  Thus we see that the fates of quantum computing and
quantum teleportation are entangled!

\section{Quantum teleportation}\sectlabel{teleport}

Recall that any attempt at measuring quantum information
disturbs it irreversibly and yields incomplete information.
This makes it impossible to transmit quantum information
through a classical channel.  Recall also that the purpose
of quantum teleportation~\cite{BBCJPW} is to circumvent this
impossibility so as to allow the faithful transmission of
quantum information between two parties, conventionally
referred to as Alice and Bob.

In~order to achieve teleportation, Alice and Bob must
share prior quantum entanglement.  This is usually explained in terms
of Einstein--Podolsky--Rosen nonlocal quantum states~\cite{EPR}
and Bell measurements, which makes the process seem very mysterious.
The~purpose of this note is to show how to achieve quantum teleportation
very simply in terms of quantum computation.  As~interesting side
product, we obtain a quantum circuit with the unusual feature that
there are points in the circuit at which the quantum information can
be completely disrupted by a measurement---or~some types of interaction
with the environment---without ill effects:
the same final result is obtained whether or not measurement takes place.
This is true despite that fact that the qubits affected by these
measurements are entangled with the other qubits carried by the
circuit, which should make these measurements even more damaging.

Of~course, the uncanny power of quantum computation draws in parts
on nonlocal effects inherent to quantum mechanics.
The~quantum teleportation circuit described in~\sect{circuit} is not really
different in principle from the original idea~\cite{BBCJPW} since
it uses quantum computation to create and measure nonlocal states.
Nevertheless it sheds new light on teleportation, at least from a
pedagogical point of view, since it makes the process completely
straightforward to anyone who believes that quantum computation is
a reasonable proposition.  Moreover, this circuit could genuinely be
used for teleportation purposes inside a quantum computer.
Finally, the surprising resilience of this circuit to measurements
performed while it is processing information
may turn out to have relevance to quantum error correction. 

\section{The basic ingredients}\sectlabel{ingredients}

As is often the case with quantum computation, we shall need two basic
ingredients: the exclusive-or gate (also known as controlled-not), which
acts on two qubits at once, and arbitrary unitary operations on single qubits.
Let~\ket{0} and~\ket{1} denote basis states for single qubits and recall
that pure states are given by linear combination of basis states
such as \mbox{$\ket{\psi}=\alpha\ket{0}+\beta\ket{1}$} where $\alpha$ and
$\beta$ are complex numbers such that \mbox{$\norm{\alpha}+\norm{\beta}=1$}.

The quantum exclusive-or (XOR), denoted as follows,
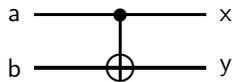
\begin{center}
\begin{picture}(180,60)
\thicklines
\put(0,40){\makebox(20,20){\sf a}}
\put(25,50){\line(1,0){130}}
\put(90,50){\circle*{10}}
\put(160,40){\makebox(20,20){\sf x}}
\put(0,0){\makebox(20,20){\sf b}}
\put(25,10){\line(1,0){130}}
\put(90,10){\circle{20}}
\put(160,0){\makebox(20,20){\sf y}}
\put(90,0){\line(0,1){50}}
\end{picture}
\end{center}
sends \ket{00} to \ket{00}, \ket{01} to \ket{01}, \ket{10} to \ket{11}
and \ket{11} to~\ket{10}.  In~other words,
{\em provided the input states at {\sf a} and {\sf b} are in basis states},
the output state at {\sf x} is the same as the input state at {\sf a},
and the output state at {\sf y} is the exclusive-or of the two input states
at {\sf a} and~{\sf b}.  This is also known as the controlled-not gate because
the state carried by the {\em control} wire ``{\sf ax}'' is not disturbed
whereas the state carried by the {\em controlled} wire ``{\sf by}'' is
flipped if and only if the state on the control wire was~\ket{1}.
Note that the classical interpretation given above no longer holds 
if the input qubits are not in basis states: it is possible for
the output state on the control wire (at~{\sf x}) to be different from its
input state (at~{\sf a}).  Moreover, the joint state of the output
qubits can be entangled even if the input qubits were not, and
vice versa.

In addition to the quantum exclusive-or, we shall need two single-qubit
rotations {\sf L} and {\sf R}, and two single-qubit conditional phase-shifts
{\sf S} and~{\sf T}\@.  Rotation {\sf L} sends \ket{0} to
\mbox{$(\ket{0}+\ket{1})/\sqrt{2}$} and \ket{1}
to \mbox{$(-\ket{0}+\ket{1})/\sqrt{2}$},
whereas {\sf R}~sends \ket{0} to
\mbox{$(\ket{0}-\ket{1})/\sqrt{2}$} and \ket{1}
to \mbox{$(\ket{0}+\ket{1})/\sqrt{2}$}.
Note that \mbox{$\mbox{\sf LR}\ket{\psi}=\mbox{\sf RL}\ket{\psi}=\ket{\psi}$}
for any qubit~\ket{\psi}.
Conditional phase-shift {\sf S} sends \ket{0} to $i\ket{0}$
and leaves \ket{1} undisturbed, whereas {\sf T} sends
\ket{0} to $-\ket{0}$ and \ket{1} to $-i\ket{1}$.
In~terms of unitary matrices, the operations are
\[
\begin{array}{lll}\multicolumn{2}{l}{
\mbox{\sf L} = {\displaystyle \frac{1}{\sqrt{2}}}
\left( \begin{array}{rr}1&-1\\1&1\end{array} \right)} &
\mbox{\sf R} = {\displaystyle \frac{1}{\sqrt{2}}}
\left( \begin{array}{rr}1&1\\-1&1\end{array} \right) \\[7mm]
\mbox{\sf S} = \left( \begin{array}{rr}i&0\\0&1\end{array} \right)
& \mbox{and} &
\mbox{\sf T} = \left( \begin{array}{rr}-1&0\\0&-i\end{array} \right)
\end{array}
\]
if $\alpha\ket{0}+\beta\ket{1}$ is represented by vector \state{\alpha}{\beta}.
Similarly the quantum exclusive-or operation is given by matrix
\[\mbox{\sf XOR} = \left(
\begin{array}{llll}1&0&0&0\\0&1&0&0\\0&0&0&1\\0&0&1&0\end{array}\right)\]
if \mbox{$\alpha\ket{00}+\beta\ket{01}+\gamma\ket{10}+\delta\ket{11}$} is
represented by the transpose of vector \mbox{$(\alpha,\beta,\gamma,\delta)$}.

\section{The teleportation circuit}\sectlabel{circuit}

Consider the following quantum circuit.  Please disregard the dashed line
for the moment.
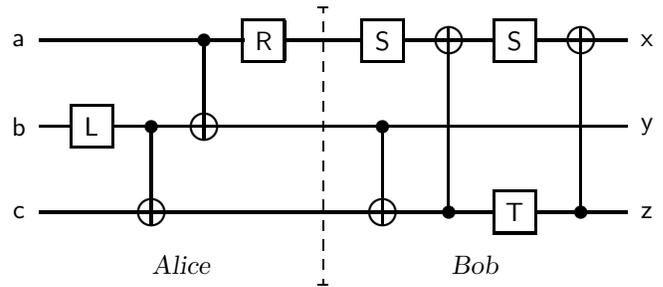
\begin{center}
\begin{picture}(495,210)(0,-40)
\thicklines
\put(0,135){\makebox(20,20){\sf a}}
\put(25,145){\line(1,0){155}}
\put(150,145){\circle*{10}}
\put(180,130){\framebox(30,30){\sf R}}
\put(210,145){\line(1,0){60}}
\put(270,130){\framebox(30,30){\sf S}}
\put(300,145){\line(1,0){70}}
\put(335,145){\circle{20}}
\put(370,130){\framebox(30,30){\sf S}}
\put(400,145){\line(1,0){70}}
\put(435,145){\circle{20}}
\put(475,135){\makebox(20,20){\sf x}}
\put(0,70){\makebox(20,20){\sf b}}
\put(25,80){\line(1,0){25}}
\put(50,65){\framebox(30,30){\sf L}}
\put(80,80){\line(1,0){390}}
\put(110,80){\circle*{10}}
\put(150,80){\circle{20}}
\put(285,80){\circle*{10}}
\put(475,70){\makebox(20,20){\sf y}}
\put(0,5){\makebox(20,20){\sf c}}
\put(25,15){\line(1,0){345}}
\put(110,15){\circle{20}}
\put(285,15){\circle{20}}
\put(335,15){\circle*{10}}
\put(370,0){\framebox(30,30){\sf T}}
\put(400,15){\line(1,0){70}}
\put(435,15){\circle*{10}}
\put(110,5){\line(0,1){75}}
\put(150,70){\line(0,1){75}}
\put(285,5){\line(0,1){75}}
\put(335,15){\line(0,1){140}}
\put(435,15){\line(0,1){140}}
\put(475,5){\makebox(20,20){\sf z}}
\thinlines
\put(240,-40){\dashbox{7.5}(0,210){}}
\put(25,-40){\makebox(215,30){\sl Alice}}
\put(240,-40){\makebox(230,30){\sl Bob}}
\end{picture}
\end{center}
Let \ket{\psi} be an arbitrary one-qubit state.  Consider
what happens if you feed \ket{\psi00} in this circuit,
that is if you set upper input {\sf a} to \ket{\psi} and both other
inputs {\sf b} and {\sf c} to~\ket{0}.  It~is a straightforward
exercise to verify that state \ket{\psi} will be transferred
to the lower output~{\sf z}, whereas both other outputs {\sf x} and {\sf y}
will come out in state \mbox{$\ket{\phi}=(\ket{0}+\ket{1})/\sqrt{2}$}.
In~other words the output will be~\ket{\phi\phi\psi}.
If~the two upper outputs are measured in the standard basis
\mbox{(\ket{0} versus \ket{1})}, two random classical bits will be obtained
in addition to quantum state \ket{\psi} on the lower \mbox{output}.

Now, let us consider the state of the system at the dashed line.
A~simple calculation shows that all three qubits are entangled.
We~should therefore be especially careful not to
disturb the system at that point.  Never\-the\-less, let us 
measure the two upper qubits, leaving the lower qubit undisturbed.
This~measurement results in two purely random classical bits $u$ and~$v$,
bearing no correlation whatsoever with the original state~\ket{\psi}.
Let~us now turn $u$ and $v$ back into quantum bits and reinject
\ket{u} and \ket{v} in the circuit immediately after the dashed line.

Needless to say that the quantum state carried at the dashed line
has been completely disrupted by this measurement-and-resend process.
We~would therefore expect this disturbance to play havoc with the
final output of the circuit.  Not~at all!  In~the end, the state
carried at {\sf xyz} is~\ket{uv\psi}.  In~other words, \ket{\psi}
is still obtained at~{\sf z} and the other two qubits, if measured,
are purely random provided we forget the measurement outcomes at the
dashed line.  Another way of seeing this phenomenon is that the
outcome of the circuit will not be altered if the state of the upper
two qubits leaks to the environment (in the standard basis)
at the dashed line.

To~turn this circuit into a quantum teleportation \mbox{device}, we need
the ability to store qubits.  Assume Alice prepares two qubits in
state \ket{0} and pushes them through the first two gates of the
circuit.
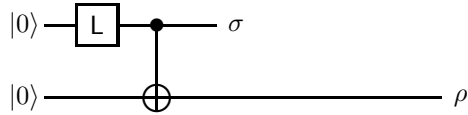
\begin{center}
\begin{picture}(325,80)(0,5)
\thicklines
\put(0,60){\makebox(20,20){\ket{0}}}
\put(25,70){\line(1,0){25}}
\put(50,55){\framebox(30,30){\sf L}}
\put(80,70){\line(1,0){75}}
\put(160,60){\makebox(20,20){$\sigma$}}
\put(110,70){\circle*{10}}
\put(0,5){\makebox(20,20){\ket{0}}}
\put(25,15){\line(1,0){300}}
\put(110,15){\circle{20}}
\put(330,5){\makebox(20,20){$\rho$}}
\put(110,5){\line(0,1){65}}
\end{picture}
\end{center}
She keeps the upper qubit $\sigma$ in quantum memory and gives the other,
$\rho$, to Bob.  [We~do not denote these qubits by kets because
they are not individual pure states: \mbox{together} they are in state
\mbox{$\Phi^{+}=(\ket{00}+\ket{11})/\sqrt{2}$}.]
At~some later time, Alice receives a mystery qubit in \mbox{unknown}
state~\ket{\psi}. In~order to teleport this qubit to Bob, she releases
$\sigma$ from her quantum memory and pushes it together with the mystery
qubit through the next two gates of the circuit.  She measures both
output wires to turn them into classical bits $u$ and~$v$.
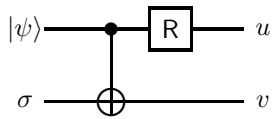
\begin{center}
\begin{picture}(200,80)(75,70)
\thicklines
\put(75,125){\makebox(20,20){\ket{\psi}}}
\put(100,135){\line(1,0){80}}
\put(150,135){\circle*{10}}
\put(180,120){\framebox(30,30){\sf R}}
\put(210,135){\line(1,0){40}}
\put(255,125){\makebox(20,20){$u$}}
\put(75,70){\makebox(20,20){$\sigma$}}
\put(100,80){\line(1,0){150}}
\put(150,80){\circle{20}}
\put(255,70){\makebox(20,20){$v$}}
\put(150,70){\line(0,1){65}}
\end{picture}
\end{center}
To complete teleportation, Alice has to communicate $u$ and $v$ to Bob
by way of a classical communication channel.  Upon reception of the
signal, Bob creates quantum states \ket{u} and \ket{v} from the
classical information received from Alice, he releases the qubit
$\rho$ he had kept in quantum memory, and he pushes all three
qubits into his part of the circuit (on~the right of the dashed line).
Finally Bob may wish to measure the two upper qubit at {\sf x} and~{\sf y}
to make sure that he gets $u$ and~$v$; otherwise something went wrong
in the teleportation apparatus.  At~this point, teleportation is
complete as Bob's output {\sf z} is in state~\ket{\psi}.
Note that this process works equally well if Alice's mystery qubit
is not in a pure state.  In~particular, Alice can teleport to Bob
entanglement with an arbitrary auxiliary system, possibly outside both
Alice's and Bob's labo\-ra\-tories.

In~practice, Bob need not use the quantum circuit shown right of
the dashed line at~all.  Instead, he may choose classically one of 4
possible rotations to apply to the qubit he had kept in quantum memory,
depending on the 2 classical bits he receives from Alice.
(This would be more in tune with the original teleportation
proposal~\cite{BBCJPW}.)  This explains the earlier claim
that quantum teleportation can be achieved at the cost of
only two quantum exclusive-ors: those of Alice.
Never\-theless, the unitary version of Bob's process
given here may be more appealing than choosing classically
among 4 courses of action if teleportation is used
inside a quantum computer.

%end \small

{\small
\begin{thebibliography}{99}
\sectlabel{references}
\let\bib\bibitem
        \bib{BBBSS}{\sc Bennett}, Charles~H., Fran\c{c}ois {\sc Bessette},
         Gilles {\sc Brassard}, Louis {\sc Salvail} and John {\sc Smolin},
        ``Experimental quantum cryptography'',
        {\it Journal of Cryptology} {\bf 5}:1 (1992), \mbox{3\,--\,28}.
\let\bib\bibitem
        \bib{BB84}{\sc Bennett}, Charles~H. and Gilles {\sc Brassard},
        ``Quantum cryptography: Public key distribution and coin tossing'',
        {\it Proceedings of IEEE International Conference on Computers,
        Systems and Signal Processing}, Bangalore, India
        (1984), \mbox{175\,--\,179}.
\let\bib\bibitem
        \bib{BBCJPW}{\sc Bennett}, Charles~H., Gilles {\sc Brassard},
        Claude {\sc Cr\'epeau}, Richard {\sc Jozsa}, Asher {\sc Peres}
        and William {\sc Wootters},
        ``Teleporting an unknown quantum state via dual classical and
        Einstein--Podolsky--Rosen channels'',
        {\it Physical Review Letters} {\bf 70}:13 (1993),
        \mbox{1895\,--\,1899}.
\let\bib\bibitem
        \bib{sciam}{\sc Bennett}, Charles~H., Gilles {\sc Brassard} and
        Artur {\sc \mbox{Ekert}},
        ``Quantum cryptography'',
        {\it Scientific American} (October 1992), \mbox{50\,--\,57}.
\let\bib\bibitem
        \bib{LNCS}{\sc Brassard}, Gilles,
        ``A quantum jump in computer science'',
        in {\it Computer Science Today}
        (Jan van Leeuwen ed.),
        Lecture Notes in Computer Science {\bf 1000},
        Springer-Verlag (1995), \mbox{1\,--\,14}.
\let\bib\bibitem
        \bib{RSA}{\sc Brassard}, Gilles,
        ``The~impending demise of RSA?'',
        {\it RSA Laboratories CryptoBytes} {\bf 1}:1 (1995), \mbox{1\,--\,4}.
\let\bib\bibitem
        \bib{deutsch}{\sc Deutsch}, David,
        ``Quantum theory, the Church--Turing principle and the
        universal quantum computer'',
        {\it Proceedings of the Royal Society, London}
        {\bf A400} (1985), \mbox{97\,--\,117}.
\let\bib\bibitem
        \bib{EPR}{\sc Einstein}, Albert, Boris {\sc Podolsky}
        and Nathan {\sc Rosen},
        ``Can~quantum-mechanical description of physical reality
        be considered complete?'',
        {\it Physical Review} {\bf 47} (1935), \mbox{777\,--\,780}.
\let\bib\bibitem
        \bib{feynman}{\sc Feynman}, Richard~P.,
        ``Quantum mechanical computers'',
        {\it Optics News} (1985);
        Reprinted in {\it Foundations of Physics} {\bf 16}:6 (1986),
        \mbox{507\,--\,531}.
\let\bib\bibitem
        \bib{hughes}{\sc Hughes}, Richard~J., G.\,G.~{\sc Luther},
        G.\,L.~{\sc Morgan}, C.\,G.~{\sc Peterson} and C.~{\sc Simmons},
        ``Quantum cryptography over underground optical fibers'',
        {\it Advances in Cryptology: Crypto~'96 Proceedings},
        Springer-Verlag (1996).
\let\bib\bibitem
        \bib{IBM}{\sc International Business Machines},
        ``Stand by: I'll teleport you some goulash'',
        {\it Scientific American} (February 1996), \mbox{0\,--\,1} ({\em sic}).
\let\bib\bibitem
        \bib{townsend}C. {\sc Marand} and Paul~D. {\sc Townsend},
        ``Quantum key distribution over distances as long as 30\,km'',
         {\it Optics Letters} {\bf 20} (1995).
\let\bib\bibitem
        \bib{gisin}{\sc Muller}, Antoine, Hugo {\sc Zbinden} and
        Nicolas {\sc Gisin},
        ``\mbox{Underwater} quantum coding'',
        {\it Nature} {\bf 378} (1995), 449.
\let\bib\bibitem
        \bib{shor}{\sc Shor}, Peter~W.,
        ``Algorithms for quantum computation: Discrete logarithms and
        factoring'',
        {\it Proceedings of the 35th Annual IEEE Symposium
        on Foundations of Computer Science} (1994), \mbox{124\,--\,134}.


\end{thebibliography}
 }
\end{document}